\documentstyle [prb,aps]{revtex}
\begin {document}
\draft
\title{Ginzburg-Landau theory of phase transitions
in quasi-one-dimensional systems}

\author{Ross H. McKenzie\cite{email}}

\address{School of Physics, University of New
South Wales, Sydney, NSW 2052, Australia}

\maketitle
\begin{abstract}
A wide range of quasi-one-dimensional materials,
consisting of weakly coupled chains,
undergo three-dimensional phase transitions
that can be described by a complex order parameter.
A Ginzburg-Landau theory is derived for such a transition.
It is shown that intrachain fluctuations in the
order parameter play a crucial role and must be treated exactly.
The effect of these fluctuations is determined
by a single dimensionless parameter.
The three-dimensional transition temperature,
the associated specific heat jump, coherence lengths,
and width of the critical region,
are computed assuming that the single chain Ginzburg-Landau
coefficients are independent of temperature.
The width of the critical region, estimated
from the Ginzburg criterion, is virtually
parameter independent, being about 5-8 per cent of
the transition temperature.\\
 \\
To appear in {\it Physical Review B,} March 1, 1995.
\end{abstract}



\section{Introduction}

\subsection{Motivation
\label{motiv}}

Crystals consisting of linear molecular chains
can have highly anisotropic electronic properties.
Many of these quasi-one-dimensional
materials undergo  finite temperature
phase transitions.  Detailed experimental
studies in the 1970's of the charge-density-wave (CDW) transition
in materials such as KCP and TTF-TCNQ
\cite{too} stimulated work on the theoretical
description of phase transitions in quasi-one-dimensional
materials \cite{lee,sca2,man,bis,die,sch}.
Unfortunately, due to the limited quality of samples
it was not possible to make a quantitative comparison
of experiment with theory. In the past decade, a whole
new range of quasi-one-dimensional
materials, many in high-purity single crystals, have been
synthesized \cite{icsm}.
Examples of new materials are the CDW material
K$_{0.3}$MoO$_3$ (blue bronze),
the Bechgaard salt (TMTSF)$_2$PF$_6$ \cite{gru},
that undergoes a spin-density-wave (SDW) transition,
 thin wires of superconducting lead \cite{sha},
and an inorganic compound CuGeO$_3$ that undergoes a
spin-Peierls transition \cite{has}.
Furthermore, high-quality experimental data
\cite{dat,bri,cor} on these crystals opens the possibility of
making a quantitative comparison of experiment with
theory. For example, recently anomalies in the specific heat,
Young's modulus, thermal expansion, and magnetic susceptibility
close to the three-dimensional CDW transition in K$_{0.3}$MoO$_3$
were all precisely measured on the {\it same} single
crystal \cite{bri}.

Fluctuations in the order parameter are extremely important
in these materials. In a strictly one-dimensional system
with short-range interactions there are
no phase transitions at finite temperature
because fluctuations in the order parameter destroy
long range order \cite{lan,sca}. Consequently in a real
quasi-one-dimensional material a finite-temperature
phase transition only occurs as a result of the
weak interchain interactions.
Hence, a complete theory must treat fluctuations along the
chain carefully and also include the interchain
interactions. On the other hand, in most materials the
three-dimensional transition temperature, $T_{3D}$,
is clearly defined and the width of the critical region
is only a few percent of $T_{3D}$.
Hence, a three-dimensional
Ginzburg-Landau theory should accurately describe the transition,
except in a narrow temperature range.

The purpose of the present paper is to provide a
complete derivation of a Ginzburg-Landau theory
describing the three-dimensional ordering transition
of a quasi-one-dimensional system with a complex
(i.e., two-component) order parameter.
Such an order parameter describes  superconductors,
charge- and spin-density waves (which are incommensurate
with the lattice), and a spin-Peierls system in a large magnetic field.
Particular attention is given to examining the
validity of the assumptions and approximations
made in previous work \cite{sca2,man,bis,die,sch}.
This work is the continuation of a larger program \cite{mck,kim,mck2}
of examining the effect that lattice fluctuations
have on electronic properties of CDW compounds.
Recently, the role of order parameter fluctuations in the blue bronze
and the Bechgaard salts was considered by Castella,
Baeriswyl, and Maki \cite{cas}.
Schulz and Bourbonnais have considered quantum fluctuations
in the phase of the order parameter in quasi-one-dimensional
superconductors \cite{sch2}.

\subsection{Overview}

The Ginzburg-Landau free energy functional $F_1[\phi]$
for a {\it single} chain with a complex order
parameter $\phi(z)$, where $z$ is the co-ordinate along
the chain, is
\begin{equation}
F_1[\phi]=\int dz \left[
a \mid\phi\mid^2 + \ b  \mid\phi\mid^4 +
 \  c \mid {\partial \phi\over \partial z}\mid ^2 \right]
   \label{aa1}
\end{equation}
In this paper the
coefficients $a$, $b$, and $c$ will be treated as
phenomological parameters. For a specific system
these coefficients can be  calculated from
microscopic theory \cite{sch,coe}.

Due to fluctuations in the order parameter
this system cannot develop long-range order at
finite temperature \cite{lan,sca}.
Furthermore, the values for the coefficients $a$, $b$, and $c$
given by the simplest microscopic theories predict that
the fluctuations are important over a temperature range
comparable to the single chain mean-field transition
temperature \cite{sch}.
To describe a finite-temperature
phase transition, a set of weakly interacting
chains are considered.  If $\phi_i(z)$ is the order parameter on the $i$-th
chain
the free energy functional for the system is
\begin{equation}
F[\phi_i(z)]=\sum_i F_1[\phi_i(z)] -
\sum_{i,j} J_{ij} \int dz {\rm Re} [\phi_i(z)^*  \phi_j(z)]
   \label{ad1}
\end{equation}
where $J_{ij}$ describes the interchain interactions.
In most of this paper it will be assumed that the interchain interaction
$J_{ij}$
is non-zero only for nearest neighbour chains
and that its value is $J_x/4$ and $J_y/4$ in the $x$ and $y$
directions respectively.
A mean-field treatment of this functional will not
give accurate results due to the large intrachain
fluctuations (see Section \ref{deriv}).
 This problem is solved by integrating
out these fluctuations to derive a new Ginzburg-Landau
functional with renormalized coefficients,
\begin{equation}
\tilde F[\Phi(x,y,z)]= {1 \over a_x a_y}\int d^3 x \left[
A \mid \Phi \mid^2
+B \mid \Phi \mid^4
+ C_x \mid {\partial \Phi \over \partial x }\mid^2
+C_y  \mid {\partial \Phi \over \partial y }\mid^2
+ C_z \mid {\partial \Phi \over \partial z }\mid^2
\right]
\label{bg1}
\end{equation}
where $a_x$ and $a_y$ are the lattice constants
perpendicular to the chains.
The new order parameter $\Phi(x,y,z)$, is proportional
to the average of
$\phi_i(z)$ over neighbouring chains (see equation (\ref{ay1})).
The three-dimensional mean-field temperature $T_{3D}$ is defined
as the temperature at which the the coefficient $A(T)$
changes sign. Close to $T_{3D}$
\begin{equation}
A=A^\prime\left({T \over T_{3D}} -1 \right).
\label{abg1}
\end{equation}

The goal of the present paper is to derive the
functional (\ref{bg1}) and find expressions for
the transition temperature $T_{3D}$ and the
coefficients $A^\prime$, $B$, $C_x$, $C_y$, and $C_z$
in terms of the interchain interaction
$J_{ij}$ and  the coefficients
$a$, $b$, and $c$ of a single chain.

The coefficients determine measurable quantities associated
with the transition such as the specific heat jump,
coherence lengths and width of the critical region.
The specific heat jump per chain at the transition is given by
\begin{equation}
\Delta C=
{( A^\prime)^2 \over 2B T_{3D}}.
\label{ac}
\end{equation}
The specific heat jump per unit volume is
$\Delta C_{3D} \equiv \Delta C/(a_x a_y).$
In the Gaussian approximation the correlation length
 parallel to the chains is given by
\begin{equation}
\xi_z=\sqrt{{C_z \over A}}=\xi_{0z}
\left({T \over T_{3D}} -1 \right)^{-1/2}
\label{ad}
\end{equation}
where $\xi_{0z}$ is the longitudinal coherence length defined as
\begin{equation}
\xi_{0z}=\sqrt{{C_z \over A^\prime}}.
\label{ae}
\end{equation}
The correlation and coherence lengths perpendicular
to the chains are given by similar expressions.

The Ginzburg criterion \cite{gin}
gives a rough estimate of the
width, $\Delta T_{3D}$, of the 3D critical region.
\begin{equation}
\Delta t_{3D} \equiv {\Delta T_{3D} \over T_{3D}}
={1 \over 32 (\pi \Delta C_{3D}
\xi_{x0} \xi_{y0} \xi_{z0})^2}
\label{bt1}
\end{equation}

This paper establishes the striking result that most of the physics is
determined by a {\it single} dimensionless parameter
\begin{equation}
\kappa  \equiv { 2 (bT)^2 \over |a|^3 c}.
\label{aat1}
\end{equation}
A brief argument is now given to show that
$\kappa$ is a measure of the fluctuations along a single chain.
The rms fluctuation in the single chain order parameter
 $<\mid\phi\mid^2>$, calculated in the Gaussian approximation
is $T/2(c|a|)^{1/2}$. The magnitude of the order parameter,
calculated in the mean-field approximation, $\phi_0$,
is given by $\phi_0^2=|a|/2b$. Hence,
$\kappa=2 (<\mid\phi\mid^2>/\phi_0^2)^2$ and so is a
measure of the importance of fluctuations.
The strength of the interchain interactions determines the
value of $\kappa$ at the
three-dimensional transition temperature.

The general approach that has been taken
previously\cite{sca2,man,die} when deriving a
Ginzburg-Landau functional for the three-dimensional
transition is to
treat the interchain interactions in the mean-field approximation
and then solve the resulting one-dimensional problem
treating the fluctuations along the chain in the
lowest-level approximation (see Section \ref{lla1} for a definition).
In this paper the interchain interactions are also treated
in the mean-field approximation
but the fluctuations along the chain are treated exactly.
This paper is confined to a mean-field analysis of the
functional (\ref{bg1}). In Section \ref{meas} it is shown that this
is justified except in a narrow temperature range very close
to $T_{3D}$. However, it should be pointed out that
one could perform a sophisticated renormalization group
analysis for the functional (\ref{bg1})
such as that due to Chen \cite{che}
and which has been recently used in the analysis of
specific heat measurements on the CDW compound
K$_{0.3}$MoO$_3$ \cite{bri}. The free energy functional
that derived here could be used as the input to such an analysis.

The outline of the paper is as follows.
Table \ref{table2} contains a summary of the symbols for
the important quantities for a single chain and for
a three-dimensional system of weakly interacting chains.
Section II describes an {\it exact} treatment
of the fluctuations in a strictly one-dimensional system.
Section \ref{deriv} contains a complete derivation of the coefficients
in the Ginzburg-Landau functional (\ref{bg1}).
In Section \ref{meas} it is assumed that the single chain coefficients
$a$, $b$, and $c$ are temperature independent.
The three-dimensional transition temperature is
calculated as a function of the interchain coupling
(Figure \ref{fig3a}).
The values derived for the coefficients $A^\prime$, $B$,
$C_x$, $C_y$, and $C_z$ are used to calculate the
specific heat jump, coherence lengths, and width of the
critical region at the three-dimensional transition
(Figures \ref{fig3} and \ref{fig4}).
The striking result is established that the width of the
critical region is fairly independent of any parameters,
$\Delta t_{3D}\simeq 0.05-0.08$.
Sections \ref{dgl} and \ref{deriv} can be omitted by readers not interested
in technical details.
The Hartree, Hartree-Fock, and lowest-level approximations
for the fluctuations that have been given
in previous work are described in the Appendix.
 It is shown that the lowest-level approximation
is only reliable for $\kappa < 10^{-3}$
(Figures \ref{fig1} and \ref{fig2}).
In Table \ref{table1} very rough estimates of $\kappa$ for various
CDW materials give $\kappa > 10^{-2}$, showing that
they are well outside the regime of
validity of the lowest-level approximation.

\section{One-dimensional Ginzburg-Landau theory}
\label{dgl}

The mean-field treatment of one-dimensional Ginzburg-Landau theory
is described before an exact treatment of
the order-parameter fluctuations is given.
For completeness the Appendix to the paper describes
various approximate treatments of fluctuations that
have been given in earlier work.

\subsection{Mean-field treatment}

Long-wavelength fluctuations associated
with the gradient term in (\ref{aa1}) are
neglected. One simply minimizes the potential
\begin{equation}
V(\phi)= a \mid\phi\mid^2 + \ b  \mid\phi\mid^4.
   \label{aaa1}
\end{equation}
The single chain mean-field transition temperature $T_0$ is
defined by the temperature at which the coefficient $a(T)$ vanishes.
Close to $T_0$ it is convenient to write
\begin{equation}
a(T)= (t -1 ) a^\prime
\label{ab1}
\end{equation}
where $t = T / T_0$.
In the mean-field approximation there is a phase transition
at $T_0$ and the magnitude of the order parameter
for $T < T_0$,
$\phi_0$, is
\begin{equation}
\phi_0^2 = { \mid a \mid \over 2 b}.
\label{aq1}
\end{equation}
The free energy per unit length of the ordered state, relative to the
disordered state is smaller by $V_0$, given by
\begin{equation}
V_0 \equiv -V[\phi_0] = { a^2 \over 4 b}.
\label{ar1}
\end{equation}
The jump in the specific heat at the transition is
\begin{equation}
\Delta C_{1D}=
{(a^\prime)^2 \over 2 \ b  \ T_0}.
\label{ac1}
\end{equation}
An important length scale is the coherence length $\xi_0$,
defined by
\begin{equation}
\xi_0=\left({c \over a^\prime}\right)^{1/2}
\label{acd1}
\end{equation}
The one-dimensional
Ginzburg criterion \cite{gin} provides an estimate of the
temperature range, $\Delta T_{1D}$, over which critical fluctuations
are important.
\begin{equation}
\Delta t_{1D} \equiv {\Delta T_{1D} \over T_0}
= \left({b  \ T_0 \over a^{\prime 3/2} c^{1/2}}
\right)^{2/3}
= {1 \over (2 \xi_0 \Delta C_{1D})^{2/3}}.
\label{acc1}
\end{equation}

There are several serious problems with a mean-field
treatment. First, it predicts a finite temperature
phase transition. Second, the simplest microscopic models
give $ \Delta t_{1D} \sim 1$ \cite{sch}, suggesting that the
neglected fluctuations
play an important role over a very broad temperature range.
An exact treatment of these fluctuations is now given.

\subsection{Exact solution
\label{exact}}

The goal is to evaluate the functional integral for the partition
function $Z$ of a system with the Ginzburg-Landau free energy
(\ref{aa1}),
\begin{equation}
Z=\int [d\phi(z)] \exp \left( -F_1[\phi(z)]/T \right)
   \label{af1}
\end{equation}
It is convenient to rescale the order parameter
$\phi(z) \to \phi(z)/\phi_0$ and write the functional
(\ref{aa1}) as
\begin{equation}
F_1[\phi]=V_0 \int dz \left[
-2 \mid\phi\mid^2 +  \mid\phi\mid^4 +
 \  {2 c \over \mid a \mid}
 \mid {\partial \phi\over \partial z}\mid ^2 \right]
   \label{aff1}
\end{equation}
assuming that $T < T_0$.
Scalapino, Sears, and Ferrell\cite{sca} showed that a transfer
matrix method could be used to reduce the
problem to that of diagonalizing a transfer matrix
Hamiltonian $H_{tr}$. This involves a cylindrically symmetric
Schr\"odinger-like equation
with a single complex degree of freedom $s=\rho e^{i \varphi}$,
\begin{equation}
H_{tr} \Psi_m =
 \left[ -\kappa \left( {\partial  ^2 \over \partial \rho^2} +
 { 1 \over \rho} {\partial \over \partial \rho}
 + { 1 \over \rho^2} {\partial  ^2 \over \partial \varphi^2} \right)
 - 2 \rho^2 + \rho^4 \right] \Psi_m(\rho,\varphi) =
 \lambda_m \Psi_m(\rho,\varphi).
   \label{ah1}
\end{equation}
The eigenvalues $\lambda_m$ only depend on the
dimensionless parameter $\kappa$.

As discussed in the introduction $\kappa$
is a measure of the size of the fluctuations in the order parameter,
and defined by equation (\ref{aat1}) \cite{kap}.
$\kappa$ {\it is central to this paper.}
It is shown in Section \ref{meas} that the value of
$\kappa$ at the three-dimensional
transition temperature determines physically important quantities
such as the specific heat jump.
 For the case of a real (i.e., one-component) order parameter
the importance of $\kappa$ was emphasized previously by
Bishop and Krumhansl \cite{bis} ($\kappa$ corresponds to
$2\mu^2$ in their paper) and Dieterich \cite{die}
($\kappa$ corresponds to $4/\alpha^2$ in his paper).

The partition function (\ref{af1}) for a system of length $L$ is
\begin{equation}
Z = \sum_m \exp( - L V_0 \lambda_m / T)
   \label{ak1}
\end{equation}
Consequently, in the thermodynamic limit, the
free energy per unit length equals $V_0 \lambda_0$.
The order parameter correlation function is
\begin{equation}
\langle \phi(z) \phi(0)^* \rangle = \phi_0^2\sum_m
\mid < \Psi_m | s | \Psi_0 > \mid ^2
\exp\left( - {|z|\over T} V_0 (\lambda_m - \lambda_0) \right)
\label{al1}
\end{equation}
which at large distances becomes
\begin{equation}
\langle \phi(z) \phi(0)^* \rangle  = \phi_0^2
\mid < \Psi_1 | s | \Psi_0 > \mid ^2
\exp\left( - {|z|\over \xi_1 }\right)
\label{am1}
\end{equation}
where $\xi_1$ is the correlation length given
by the separation of the lowest two eigenvalues of the
transfer matrix Hamiltonian
\begin{equation}
\xi_1 = { T \over V_0 (\lambda_1 - \lambda_0)}
= \left({c \over a}\right)^{1/2}
 (8\kappa)^{1/2}
 {1 \over (\lambda_1 - \lambda_0)(\kappa)}.
\label{an1}
\end{equation}
The Fourier transform of the correlation function
defines the static linear susceptibility
\begin{equation}
\chi_1(q)= {1 \over T} \int dz e^{-iqz} \langle \phi(z) \phi(0)^* \rangle.
\label{all1}
\end{equation}
For small $q$ it is of the form
\begin{equation}
\chi_1(q)= { \chi_1(0) \over 1 + (q \xi_1)^2 }
\label{alm1}
\end{equation}
where
\begin{equation}
\chi_1(0)=  {2 \phi_0^2 \over V_0 }\sum_m
{\mid < \Psi_m | s | \Psi_0 > \mid ^2 \over
 \lambda_m - \lambda_0 }
 \equiv {4  \over \mid a \mid } f(\kappa)
\label{aln1}
\end{equation}
and a dimensionless function,
 $f(\kappa)$, that only depends on
the fluctuation parameter $\kappa$ has been introduced.

Due to the cylindrical symmetry of the potential $V(s)$
the eigenfunctions can be written in the form
\begin{equation}
\Psi_{n,\ell}(\rho,\varphi)= {u_{n,\ell}(\rho) \over
\sqrt{2 \pi \rho} } \exp(i \ell \varphi)
\ \ \ \ \ \ \ell=0, \pm 1, \pm 2,  ...
\label{ap1}
\end{equation}
Equation (\ref{ah1}) reduces to
the one-dimensional Schr\"odinger type equation
\begin{equation}
\left[ - \kappa {  d^2 \over  d \rho^2}
  + { \kappa (\ell^2 - {1 \over 4})\over \rho^2} -2 \rho^2 + \rho^4
\right]  u_{n,\ell}(\rho)=
\lambda_{n,\ell} u_{n,\ell}(\rho).
\label{ass1}
\end{equation}
In order for the system wave function to be finite at the
origin the radial wave function satisfies the boundary condition
\begin{equation}
 u_{n,\ell}(0)=0.
\label{ast1}
\end{equation}
Normalization of the wave function requires that
\begin{equation}
\int_0^{\infty} \mid u_{n,\ell}(\rho)\mid^2 d\rho=1
\label{asu1}
\end{equation}

The integrals over the angular variable $\varphi$
in the matrix elements in the susceptibility (\ref{aln1})
can be performed explicitly.
The function $f(\kappa)$ in the susceptibility (\ref{aln1})
then reduces to
\begin{equation}
f(\kappa)=  \sum_n {\mid < u_{n,1} | \rho | u_{0,0} > \mid ^2 \over
 \lambda_{n,1} - \lambda_{0,0}}.
\label{aqq1}
\end{equation}
This expression has been evaluated numerically
by solving the eigenvalue equation (\ref{ass1}) numerically
 \cite{koo}.  The results are shown in Figure \ref{fig1}.
To check the numerical results use was made  of the sum rule
\begin{equation}
{1 \over 2 \kappa} \sum_n
( \lambda_{n,1} - \lambda_{0,0})
\mid < u_{n,1} | \rho | u_{0,0} > \mid ^2  =1
\label{aqr1}
\end{equation}
which can be derived using standard arguments \cite{mer}.

It is useful to consider the effect of
an external potential $\Phi$ on the system.
Let $\lambda(\Phi)$ be the energy of the lowest eigenstate
of the transfer operator $H_{tr} + s \Phi^* + s^* \Phi$,
where $H_{tr}$ is given by (\ref{ah1}).
Second-order perturbation theory \cite{zim} shows that
the static linear susceptibility
$\chi_1(0)$  is
 given by
\begin{equation}
\chi_1(0)= - {\phi_0^2 \over V_0} { \partial^2 \lambda(\Phi)
  \over  \partial\Phi \partial\Phi^* } \Bigg|_{\Phi=0}.
\label{aqrr}
\end{equation}
The third-order nonlinear susceptibility is
\begin{equation}
\chi_3= - {\phi_0^4 \over 4 V_0^3} { \partial^4 \lambda(\Phi)
  \over  \partial\Phi^2 {\partial\Phi^*}^2 } \Bigg|_{\Phi=0}.
\label{aqr3}
\end{equation}
Expanding the Brillouin-Wigner perturbation theory
expression for $\lambda(\Phi)$ \cite{zim}
to fourth-order in $|\Phi|$ gives

\begin{eqnarray}
\chi_3&=& { 4 \phi_0^4 \over V_0^3} \sum_{m,n,p \neq 0}
{< \Psi_0 | s | \Psi_m > < \Psi_m | s^* | \Psi_n > < \Psi_n | s | \Psi_p >
< \Psi_p | s^* | \Psi_0 > \over
 (\lambda_m - \lambda_0) (\lambda_n - \lambda_0) (\lambda_p - \lambda_0)}
 \nonumber \\
&+& { 2 \phi_0^4 \over V_0^3} \sum_{m,n,p \neq 0}
{< \Psi_0 | s | \Psi_m > < \Psi_m | s | \Psi_n > < \Psi_n | s^* | \Psi_p >
< \Psi_p | s^* | \Psi_0 > \over
 (\lambda_m - \lambda_0) (\lambda_n - \lambda_0) (\lambda_p - \lambda_0)}
 \nonumber \\
 &-& {2 \phi_0^2 \over V_0^2}\chi_1(0)
 \sum_m {\mid < \Psi_m | s | \Psi_0 > \mid ^2
 \over (\lambda_m - \lambda_0)^2 }  \equiv -{ 16 b \over a^4} g(\kappa)
\label{bd1}
\end{eqnarray}
As for the linear susceptibility the above expression
can be simplified by using the factorization of
the wavefunction (\ref{ap1}) and performing the integrals
over the angular variable $\varphi$ in the matrix elements.
The dimensionless function $g(\kappa)$, defined in (\ref{bd1}),
 only depends on the
fluctuation parameter $\kappa$ and is given by
\begin{eqnarray}
-g(\kappa)
 &=& 4\sum_{n \neq 0,m,p}
{< u_{0,0} | \rho |u_{m,1} > < u_{m,1} | \rho |u_{n,0} >
 < u_{n,0} | \rho |u_{p,1} > < u_{p,1} | \rho |u_{0,0} >
\over (\lambda_{m,1} - \lambda_{0,0})
       (\lambda_{n,0} - \lambda_{0,0}) (\lambda_{p,1} - \lambda_{0,0}) }
 \nonumber \\
 &+& 2 \sum_{m,n,p}
{< u_{0,0} | \rho |u_{m,1} > < u_{m,1} | \rho |u_{n,2} >
 < u_{n,2} | \rho |u_{p,1} > < u_{p,1} | \rho |u_{0,0} >
\over (\lambda_{m,1} - \lambda_{0,0})
       (\lambda_{n,2} - \lambda_{0,0}) (\lambda_{p,1} - \lambda_{0,0}) }
 \nonumber \\
 &-& 4f(\kappa) \sum_n
 {\mid < u_{n,1} | \rho | u_{0,0} > \mid ^2 \over
  (\lambda_{n,1} - \lambda_{0,0})^2}.
\label{bdd1}
\end{eqnarray}
The functions $f(\kappa)$ and $g(\kappa)$ will appear in
the expressions for the coefficients of the
Ginzburg-Landau functional for the three-dimensional
transition.

In summary, the free energy per unit length in the
presence of an external field $\Phi$, $\tilde{F_1}[\Phi]$,
is, to fourth-order in $|\Phi|$ ,
\begin{equation}
\tilde{F_1}[\Phi]= V_0 \lambda(0) -
\chi_1(0) \mid \Phi \mid^2 - \chi_3 \mid \Phi \mid^4.
\label{xx1}
\end{equation}
If the external field is slowly varying in space $\Phi(z)$
then the quadratic term is replaced with
\begin{eqnarray}
\Phi(z)^* \int d z' \chi_1(z-z') \Phi(z')
&=& \Phi(z)^* \int {d q \over 2 \pi} \ \chi_1(q) \Phi(q) e^{iqz}
\nonumber \\
&=&\Phi(z)^* \int {d q \over 2 \pi} \ \chi_1(0) \left(1 - q^2 \xi_1^2 \right)
\Phi(q) e^{iqz}
\nonumber \\
&=& \chi_1(0) \Phi(z)^* \left( \Phi(z)
 + \xi_1^2 {\partial^2 \Phi(z) \over \partial z^2} \right)
\label{xx2}
\end{eqnarray}
where the long-wavelength expression (\ref{alm1}) has been used.

\section{Derivation of a three-dimensional Ginzburg-Landau
functional}
\label{deriv}

In this section a new three-dimensional Ginzburg-Landau
free energy functional is derived by integrating out
the order parameter fluctuations along the chain exactly
and treating the interchain interactions in the
mean-field approximation. Since this treatment (over)emphasizes
the one-dimensional fluctuations its regime of
validity is considered briefly in the Conclusions section.
Before proceeding with the derivation a rough estimate is made
of how small the interchain interactions must be in order for
a mean-field treatment of the anisotropic functional (\ref{ad1})
to be invalid. Taking the continuum limit of (\ref{ad1})
and using the three-dimensional Ginzburg criterion\cite{gin}
gives
\begin{equation}
\Delta t_{3D} \simeq {2 \over \pi^2} (\Delta t_{1D})^3
 { (a^\prime)^2 \over J_x J_y}.
\label{gw1}
\end{equation}
Hence, if $\Delta t_{1D} \sim 1$, as predicted by
the simplest microscopic theories \cite{sch}, and if $J_x \sim J_y < 0.3
a^\prime$
(the regime considered in this paper) then
$\Delta t_{3D} > 1$ and a mean-field analysis will give poor results.

The treatment given here is similar to the derivation of the Ginzburg-Landau
functional for the Ising model by Negele and Orland \cite{neg}.
The partition function for a set of chains is
\begin{equation}
Z=\int \Pi_i [d\phi_i(z)] \exp \left( -F[\phi_i(z)]/T \right)
\label{aw1}
\end{equation}
where the free energy functional $F[\phi_i(z)]$ is given by (\ref{ad1}).
The interaction term is replaced with the
following Gaussian integral
\begin{eqnarray}
&\exp&\left({1 \over T}\sum_{i,j} J_{ij} \int dz
\ {\rm Re}[\phi_i(z)^*  \phi_j(z)]
\right) \nonumber \\
&=& A \int \Pi_i [d\Phi_i(z)]
\exp\left(- T \sum_{i,j} J_{ij}^{-1} \int dz
\ {\rm Re}[\Phi_i(z)^*  \Phi_j(z)]
 + \sum_i \int dz \ {\rm Re} [\phi_i(z)^*  \Phi_i(z)]\right)
\label{ax1}
\end{eqnarray}
where $A$ is a normalization constant.
The new field $\Phi_i(z)$ will be used below as the
order parameter to describe the three-dimensional transition.
It follows from (\ref{aw1}) and (\ref{ax1}) that
\begin{equation}
\langle \Phi_i(z) \rangle = {1 \over T}
\sum_j J_{ij} \langle \phi_j(z) \rangle
\label{ay1}
\end{equation}
Hence, the new order parameter is the average of the
single chain order parameter $\phi_j(z)$ over neighbouring
chains.  The new expression for the partition function has the
desirable feature that there are no interchain interactions
involving the order parameter $\phi_i(z)$.
Consequently, the results of Section \ref{exact} can
be used to integrate exactly over the fluctuations in
$\phi_i(z)$ along each chain.
The partition function is now a
functional integral over the new order parameter $\Phi_i(z)$
which has a free energy functional $\tilde{F}[\Phi_i(z)]$,
\begin{equation}
Z=\int \Pi_i [d\Phi_i(z)] \exp \left( -\tilde{F}[\Phi_i(z)]/T \right)
\label{az1}
\end{equation}
where
\begin{equation}
\tilde{F}[\Phi_i(z)]= \sum_i \tilde{F_1}[\Phi_i(z)]
+ T \sum_{i,j} J_{ij}^{-1} \int dz {\rm  Re}[\Phi_i(z)^*\Phi_j(z)]
\label{ba1}
\end{equation}
and $\tilde{F_1}[\Phi(z)]$ is the free energy of a single chain in
the presence of an external field $\Phi(z)$,
\begin{equation}
\exp\left( -\tilde{F_1}[\Phi(z)]/T \right)
= \int [d\phi(z)] \exp\left( - {F_1[\phi(z)] \over T}
+ \sum_i \int dz {\rm Re} [\phi(z)^*  \Phi(z)]
\right).
\label{bb1}
\end{equation}
Up to this point the analysis is exact.
A mean-field analysis of the
functional integral (\ref{az1}) will now be given. In the next Section
it is shown that this
is justified except in a narrow temperature range very close
to $T_{3D}$. However, it should be stressed that
one could perform a sophisticated renormalization group
analysis of (\ref{az1}).
The free energy functional
that derived here could be used as the input to such an analysis.

Close to $T_{3D}$, the three-dimensional transition temperature
the new order parameter
$\Phi_i(z)$ will be small and slowly varying in space.
Hence $\tilde{F_1}[\Phi(z)]$ can be evaluated in a perturbation expansion
in $\Phi(z)$ and a gradient expansion.

Use is now made of the results from Section \ref{exact}.
Let $\lambda(\Phi)$ be the energy of the lowest eigenstate
of the transfer operator $H_{tr} + s \Phi^* + s^* \Phi$
where $H_{tr}$ is given by (\ref{ah1}).
The free energy per unit length of a single chain in the presence of a
slowly varying field $\Phi(z)$ is
\begin{equation}
\tilde{F_1}[\Phi(z)]= V_0 \lambda(0) -
\chi_1(0) \left( \mid \Phi \mid^2 -
\xi_1^2 \mid {\partial \Phi \over \partial z }\mid^2 \right)
 - \chi_3 \mid \Phi \mid^4
\end{equation}
where $\chi_1(0)$ is the static linear susceptibility of
the system given by (\ref{aln1}) and $\chi_3$ is the third-order
nonlinear susceptibility given by (\ref{bd1}).

Near the three-dimensional transition temperature
the continuum limit perpendicular to
the chains ($\Phi_i(z) \to \Phi(x,y,z)$)
can be taken \cite{neg2}.
Assume that the interchain interaction $J_{ij}$
is non-zero only for nearest neighbour chains
and that its value is $J_x/4$ and $J_y/4$ in the $x$ and $y$
directions respectively. Then the interchain interaction
term in (\ref{ba1}) becomes
\begin{equation}
\sum_{i,j} J_{ij}^{-1} \int dz {\rm Re} [\Phi_i(z)^*  \Phi_j(z)]
={1 \over J a_x a_y} \int dx dy \left( \mid \Phi \mid^2 +
{a_x^2 J_x\over 4 J}\mid {\partial \Phi \over \partial x }\mid^2
+ {a_y^2 J_y\over 4 J} \mid {\partial \Phi \over \partial y }\mid^2
\right)
\label{bf1}
\end{equation}
where $a_x$ and $a_y$ are the lattice constants perpendicular
to the chains and $J \equiv {1\over 2}(J_x + J_y)$.
The final free energy functional (\ref{ba1}) is of the form
(\ref{bg1}) with coefficients
\begin{equation}
A={1 \over J} - \chi_1(0)
\label{bg1a}
\end{equation}
\begin{equation}
B= - \chi_3
\label{bg1b}
\end{equation}
\begin{equation}
C_x={a_x^2 J_x\over 4 J^2}
\label{bg1c1}
\end{equation}
\begin{equation}
C_y={a_y^2 J_y\over 4 J^2}
\label{bg1c2}
\end{equation}
\begin{equation}
C_z=\chi_1(0) \xi_1^2.
\label{bg1c3}
\end{equation}
The three-dimensional mean-field temperature $T_{3D}$ is defined by
the vanishing of the coefficient $A$,
\begin{equation}
1= J \chi_1(0)
\label{bh1}
\end{equation}
where the right hand side depends on temperature.
This equation can be written in the dimensionless form
\begin{equation}
{J \over |a|}= {1 \over 4 f(\kappa(T_{3D}))}.
\label{bhh1}
\end{equation}
The relationship between the interchain coupling $J$
and the value of $\kappa$ at the three-dimensional transition
defined by this equation is shown in Figure \ref{fig3a}.

\section{Measurable quantities at the three-dimensional transition}
\label{meas}

{\it It will now be assumed that the coefficients}
$a$, $b$, {\it and} $c$ {\it are temperature independent
and } $T_{3D} \ll T _0$, {\it so} $ |a| \simeq a^\prime$.
Consequently, the only temperature dependence in the
fluctuation parameter $\kappa$ is the factor of $T^2$
(compare equation (\ref{aat1})).
Scalapino, Imry and Pincus \cite{sca2}, Manneville \cite{man},
and Dieterich \cite{die} also made this assumption.
It is an open question as to how realistic this assumption
is for different microscopic models of the CDW transition.

First, this assumption allows us to
rewrite $\kappa$ in terms of the
width of the one-dimensional critical region $\Delta t_{1D}$ \cite{kap}
and evaluate the  three-dimensional transition temperature
as a function of the interchain coupling (Figure \ref{fig3a}).
Roughly
\begin{equation}
\left({T_{3D} \over T_0 }\right)^2
\sim {1.5 J \over |a| \Delta t_{1D}^3}.
\label{bhx1}
\end{equation}
This is a useful relation because it gives a criterion
\begin{equation}
J \ll |a| \Delta t_{1D}^3
\label{bhy1}
\end{equation}
 for how weak the interchain coupling must be in order
for the three-dimensional transition temperature
to be substantially less than the single-chain
mean-field temperature $T_0$.

The coefficient $A^\prime$ (which determines the specific
heat jump and the coherence lengths) depends on the
$\kappa$ dependence of $f(\kappa)$ and the $T$ dependence of
the fluctuation parameter $\kappa$
\begin{eqnarray}
A^\prime &=&
-{1 \over J} {d \ln \chi_1(0) \over d \ln T} \Bigg|_{T=T_{3D}}
\nonumber \\
&=& {1 \over J} \left[{d \ln |a| \over d \ln T}
- {d \ln f(\kappa) \over d \ln \kappa} {d \ln \kappa \over d \ln T}
\right] \Bigg|_{T=T_{3D}}.
\label{bhi1}
\end{eqnarray}
Assuming that the coefficients $a$, $b$, and $c$
are temperture independent (\ref{bhi1}) reduces to
\begin{equation}
A^\prime = -{2 \over J} {d \ln f(\kappa) \over d \ln \kappa}
\Bigg|_{T=T_{3D}}
\label{bhj1}
\end{equation}

The specific heat jump, coherence lengths,
and critical region width can now be calculated.
It turns out that they can each be written in a
dimensionless form that only depends on
the value of the fluctuation parameter $\kappa$
at the three-dimensional transition.
The specific heat jump per chain given by (\ref{ac}),
 (\ref{bg1b}) and (\ref{bhi1}) is
\begin{equation}
\Delta C  = { \sqrt{8} \kappa^{3/2} \over \xi_0 \ g(\kappa)}
\left( {d f(\kappa) \over d \kappa }\right)^2
\label{bhk1}
\end{equation}
where $\xi_0$ is the single chain coherence length
given by (\ref{acd1}).
Hence the specific heat jump only depends on two
quantities, $\xi_0$ and $\kappa$ (or equivalently
the interchain coupling) (see Figure \ref{fig3}).
Note that as the coherence
length decreases the specific heat jump increases.
For moderate interchain coupling ($J > 0.05 |a|$)
the specific heat jump becomes independent of
the interchain coupling. The origin of this independence is not clear.

The longitudinal coherence length is
\begin{equation}
{\xi_{0z} \over \xi_0}=
{2 \over \lambda_{0,1}(\kappa) - \lambda_{0,0}(\kappa)}
\left(- {d \ln f(\kappa) \over d \kappa }\right)^{-1/2}.
\label{bhk3}
\end{equation}
The dependence of the longitudinal coherence length
on the interchain coupling is  shown in Figure \ref{fig3}.

Quantitative comparison of these expressions
for the specific heat jump
and longitudinal coherence length
with experimental data is not possible without
accurate values of $\xi_0$ from microscopic theory.
However, for all interchain coupling strengths the product of the specific heat
jump
and the longitudinal coherence length, $\Delta C \xi_{0z}$,
depends only roughly on the interchain coupling
(see the dashed curve in Figure \ref{fig3}).
Since this quantity is dimensionless it  can be compared with experiment.
If $\Delta C_{3D}$ is the specific heat jump per unit volume then
\begin{equation}
{\Delta C_{3D} a_x a_y \xi_{0z} \over k_B} \simeq 1.2 - 2.6.
\label{bhk1b}
\end{equation}
For the blue bronze K$_{0.3}$MoO$_3$ experiment \cite{bri} gives
$\Delta C_{3D}/k_B =  (2.6 \pm 0.3 ) \times 10^{-3} \AA^{-3}$,
 $\xi_{0z} = 15 \pm 3 \AA$ \cite{dat}, and
$a_x a_y = 16 \AA^2$ which
gives $\Delta C_{3D} a_x a_y \xi_{0z}/k_B = 0.6 \pm 0.2 $.

The transverse coherence length is
\begin{equation}
{\xi_{0x} \over a_x} =
\left({J_x \over J}\right)^{1/2}
\left(- 8 {d \ln f(\kappa) \over d \ln \kappa }\right)^{-1/2}.
\label{bhk4}
\end{equation}
The dependence of the right hand side on $\kappa$ is
fairly weak.  Consequently the transverse coherence
length depends only weakly on the interchain coupling
(Figure \ref{fig4}).
This result is somewhat counterintuitive: it
might be expected that as the interchain coupling
increases the transverse coherence length increases.
However, $\xi_{0x}$ is not just determined by the
coefficient $C_x$ but also by the coefficient $A^\prime$
($\xi_{0x}=(C_x/A^\prime)^{1/2}$).
It turns out that both these quantities have roughly the
same dependence on $J$.

A crossover temperature, $T_x$,
can be defined
at which correlations between chains become weak,
by $\xi_x(T_x)=a_x$ \cite{die,sch}.
 Hence,
\begin{equation}
\Delta t_x \equiv {T_x \over T_{3D}} -1
  = \left({\xi_{0x} \over a_x}\right)^2
\label{bhk5}
\end{equation}
and it can be seen from Figure \ref{fig4} that for a tetragonal crystal
$\Delta t_x \simeq 0.1 $
virtually independent of any parameters.
This value is comparable to the value obtained for
by Schulz\cite{sch} for a slightly different model
involving only fluctuations in the phase of the order parameter.
This narrow crossover region is consistent with
estimates for K$_{0.3}$MoO$_3$
based on X-ray scattering\cite{dat}.
It may be possible to argue that this narrow
crossover region, which is comparable to the width of the
critical region estimated below, is consistent with the
neglect of interchain fluctuations in the
derivation in Section III.

The width of the critical region given by the Ginzburg criterion
(\ref{bt1}) is
\begin{equation}
\Delta t_{3D}= {1 \over 16 \pi^2} {J^2 \over J_x J_y}
{ g(\kappa)^2 (\lambda_{0,1}(\kappa) - \lambda_{0,0}(\kappa))^2 \over
- \kappa f(\kappa)^3 {d f(\kappa) \over d \kappa} }
\label{bhl1}
\end{equation}
Note that for a tetragonal crystal ($J_x=J_y=J$)
this depends solely on the fluctuation parameter $\kappa$.
Figure \ref{fig4} shows the resulting dependence of
$\Delta t_{3D}$ on the interchain coupling.
It is striking that the width of the critical
region is independent of any parameters,
 $\Delta t_{3D} \sim 0.05-0.08.$
As for the transverse coherence length this result is
somewhat surprising since intuitively it might be expected that
as the the interchain coupling increases the system
becomes less one dimensional and fluctuations decrease.
However, this independence is result of the
transverse coherence lengths being only weakly
dependent on the interchain coupling.
The values obtained for $\Delta t_{3D}$
 are comparable to the value obtained for
by Schulz\cite{sch} for a slightly different model.
It is interesting that even if the
width of the one-dimensional critical region is large,
say of order $T_0$ (i.e., $\Delta T_{1D} \sim T_0$),
the actual observable critical region will be much smaller
($\Delta T_{3D} \ll  T_{3D} \ll T_0$).
{\it This small width of the critical region is important
because it shows that a  mean-field treatment of
the three-dimensional
Ginzburg-Landau functional} (\protect\ref{bg1})
{\it is justified except in
a narrow temperature region close to }$T_{3D}$.
Note that this result is based on the assumption that
the temperature dependence of the single chain
parameters $a$, $b$, and $c$ can be neglected.

The value of $\Delta t_{3D} \sim 0.05-0.08$
can be compared to experimental results.
The analysis  in Reference \cite{hau} and the
data of Reference \cite{bri} suggests
that $\Delta t_{3D} \sim 0.01$ for the CDW transtion
in K$_{0.3}$MoO$_3$.
The data of Reference \cite{cor} suggests
that $\Delta t_{3D} \sim 0.01$ for the SDW transtion
in (TMTSF)$_2$PF$_6$.
It should be stressed that these are very rough
estimates based on the Ginzburg criterion \cite{gin}.

{\it Rough estimate of $\kappa$ in several CDW materials.}
Since $\kappa $ is such an important parameter
it is worthwhile making a very rough estimate of
its value in various materials.
It turns out that most materials lie outside the regime of
validity of the lowest-level approximation,
discussed in the Appendix.

The longitudinal coherence length given by (\ref{br1})
can be used to estimate $\kappa$.
Equation (\ref{an1}) and the lowest-level approximation,
(\ref{au1}), can be used to provide
a rough estimate of $\xi_1$,
\begin{equation}
\xi_1 \simeq \xi_0
 \left({8 \over \kappa}  \right)^{1/2}
\label{bv1}
\end{equation}
where $\xi_0$ is the single chain coherence length given
by (\ref{acd1}).
Hence, $\kappa$ is roughly given by
\begin{equation}
\kappa \sim {4 \xi_0^2 \over \xi_{z0}^2}.
\label{bw1}
\end{equation}
$\xi_{z0}$ can be estimated from X-ray and neutron
scattering experiments (Table \ref{table1}).
 Deriving an accurate value of $\xi_0$ from
microscopic theory is a subtle matter which
will not be discussed here.
For purposes of this paper it is sufficient to use the
results of the simplest microscopic theory \cite{sch}.
In that case
\begin{equation}
\xi_0 \simeq {0.23 \hbar v_F \over \Delta(0)}
\label{bx1}
\end{equation}
where $v_F$ is the Fermi velocity and $\Delta(0)$ is the
zero-temperature energy gap.
Table \ref{table1}  lists estimates of $\xi_{z0}$,
$v_F$, and $\Delta(0)$ for several materials
that undergo a three-dimensional CDW ordering transition.
These values are used in equation (\ref{bw1}) to
give an order of magnitude estimate for $\kappa$ in each material.
In all the materials $\kappa > 10^{-2}$ suggesting that the
lowest-level approximation will give poor results.
Given the experimental uncertainty in the parameters discussed
above it is hoped that this discussion will generate
more precise studies of this question.

\section{Conclusions}

The main points of this paper are the following.
(1) To derive a reliable Ginzburg-Landau functional
to describe three-dimensional ordering transitions
in quasi-one-dimensional materials one must give a
careful treatment of the large intrachain
order-parameter fluctuations.
(2) Most physical properties are determined by a {\it single}
dimensionless parameter $\kappa$, defined in equation
(\ref{aat1}), which is a measure of the size of the
intrachain fluctuations.
(3) Commonly studied charge-density-wave materials
lie outside the regime of the lowest-level approximation
and so their accurate description requires an {\it exact}
treatment of the intrachain fluctuations.
(4) If the single chain Ginzburg-Landau coefficients
are assumed to be temperature independent then
the width of the three-dimensional critical region,
estimated by the Ginzburg criterion, is virtually
independent of any parameters, $\Delta t_{3D} \sim 0.05-0.08$.

This paper leaves a number of open questions and
opportunities for future work.
(a) It needs to be established whether the temperature
dependence of the single chain Ginzburg-Landau coefficients
$a$, $b$, and $c$ is important in realistic microscopic models.
If so how are the results of Figures \ref{fig3a}, \ref{fig3}, and
\ref{fig4} modified?
(b) Quantitative comparison of the results here with measurable
quantities such as the specific heat jump and longitudinal
coherence length requires values of the single chain coherence
length $\xi_0$ from microscopic theory.
(c) For moderate interchain coupling  the specific heat jump is
given by
$\Delta C \xi_0 \simeq 0.7 $, independent of the strength of
the interchain coupling. Is there a simple physical argument
that can justify this simple result?
(d) This paper treats the intrachain fluctuations exactly and
the interchain interactions in the mean-field approximation.
This neglect of the interchain fluctuations
 is presumably reasonable if the system is sufficiently anisotropic.
It is desirable to have a quantitative criterion
that defines the validity of this treatment.
Furthermore, presumably when the interchain interactions
become sufficiently strong the system is
better described by an anisotropic XY model in which the
fluctuations in all three directions are treated on an
equal footing.
It would be nice to have a description of the crossover
from a transition at $T_{3D} \ll T_0$ (ensured by
the criterion (\ref{bhy1}))
to the case $T_{3D} \sim T_0$ for strong interchain coupling.
It is not clear how to attack this problem within any particular
approximation scheme. It may be necessary to use a
numerically intensive technique such as
series expansions or a Monte Carlo simulation.
Hopefully this paper will stimulate such a study.

\acknowledgements

I have benefitted from numerous discussions with J. W. Wilkins.
This work was stimulated by discussions with J. W. Brill.
I am grateful to him for showing me his group's data
prior to publication.
I thank J. W. Brill, G. Mozurkewich and  J. W. Wilkins
for helpful comments on the manuscript.
Some computational assistance was provided by M. C. B. Ashley.
Some of this work was performed at The Ohio State University
where it was supported by the U.S. Department of Energy,
Basic Energy Sciences, Division of Materials Science
and the OSU Center for Materials Research.
Work at UNSW was supported by a Queen Elizabeth II Fellowship
from the Australian Research Council.

\begin{figure}
\caption{
Dependence of the three-dimensional transition temperature $T_{3D}$
and the fluctuation parameter $\kappa$
on the interchain coupling $J$.
It has been assumed that the single chain Ginzburg-Landau parameters
$a$, $b$, and $c$ are independent of temperature.
The reduction of $T_{3D}$ below the mean-field
transition temperature $T_0$ for a single chain also
depends on the width of the one-dimensional fluctuation
region $\Delta t_{1D}$.
\label{fig3a}}
\end{figure}

\begin{figure}
\caption{
Dependence of the specific jump per chain $\Delta C$
and the longitudinal coherence length $\xi_{0z}$
on the interchain coupling $J$.
It has been assumed that the single chain Ginzburg-Landau parameters
$a$, $b$, and $c$ are independent of temperature.
The longitudinal coherence length  is normalized to the single chain
coherence length $\xi_0$ defined by (\protect\ref{acd1}).
Note that for moderate interchain coupling ($J > 0.05 |a|$)
the specific heat jump is determined solely by
the the single chain coherence length $\xi_0$.
\label{fig3}}
\end{figure}

\begin{figure}
\caption{
Very weak dependence of the width of the critical region
for the three dimensional transition
 and the transverse coherence length
on the interchain coupling.
The width of the critical region $\Delta t_{3D}$
is calculated from the Ginzburg criterion \protect\cite{gin}.
The transverse coherence length $\xi_{0x}$ is normalized to the
lattice constant $a_x$.
The results shown are for a tetragonal crystal. For an
orthorhombic crystal the vertical scale is changed
by a factor depending on the anisotropy
(see equations (\protect\ref{bhk4}) and (\protect\ref{bhl1})).
This figure establishes the striking result that
the width of the critical region is virtually
parameter independent.
\label{fig4}}
\end{figure}

\begin{figure}
\caption{Failure of different approximation schemes for
treating the effects of fluctutations in a single chain.
The dependence of the order parameter susceptibility
function $f(\kappa)$ (defined in equation (\protect\ref{aln1}))
 on the fluctuation parameter $\kappa$ (defined
in equation (\protect\ref{aat1})) is shown.
The solid curve is the exact result
obtained by numerically solving the eigenvalue equation (\protect\ref{ass1}).
The dotted curve is the result, $f(\kappa)=1/\kappa$,
of the lowest-level approximation (LLA) \protect\cite{man}
which has been used in previous treatments of
\protect\cite{sca2,man,die} of quasi-one-dimensional phase transitions.
The dashed curves
are the results of the Hartree \protect\cite{mas} and
Hartree-Fock \protect\cite{tuc} approximations.
\label{fig1}}
\end{figure}

\begin{figure}
\caption{
Deviation of the lowest-level approximation from
the exact results.
 The linear susceptibility
$f(\kappa)$,  the third-order susceptibility $g(\kappa)$
(defined in equation (\protect\ref{bd1})), and the derivative
$ {d \ln f \over d \ln \kappa}$ (which determines the coefficient
$A^\prime$ (see equation (\protect\ref{bhj1})), all normalized to
their values in the lowest-level approximation, are shown
as a function of  the fluctuation parameter $\kappa$.
Note that the horizontal scale is logarithmnic.
Clearly the lowest-level approximation is only
quantitatively reliable for $\kappa < 10^{-3}$.
In Table \protect\ref{table1} it is estimated that typical materials have
$10^{-2} < \kappa < 1$. In this range the
lowest-level approximation will predict a specific heat
jump up to an order of magnitude smaller than the
exact results.
\label{fig2}}
\end{figure}

\begin{table}
\caption{
Summary of symbols for the important quantities for
a single chain and a three-dimensional (3D)
system of weakly coupled chains. The items in
parentheses refer to relevant equations and figures.
}
\begin{tabular}{lcc}
Quantity & Single chain & 3D\\
\tableline
Order parameter &$\phi(z)$ & $\Phi(x,y,z)$ (\protect\ref{ay1})\\
\vtop{\baselineskip=6 pt \halign{#\hfil \cr
Ginzburg-Landau \cr free energy\cr}}& (\protect\ref{aa1})
 & (\protect\ref{bg1})\\
\vtop{\baselineskip=6 pt \halign{#\hfil \cr
Mean-field \cr transition temperature\cr}}&
$T_0$ & $T_{3D} \ll T_0$ [Fig. \protect\ref{fig3a}]\\
\vtop{\baselineskip=6 pt \halign{#\hfil \cr
Ginzburg-Landau \cr coefficients\cr}}&$a$, $b$, $c$
 & $A$, $B$, $C_x$, $C_y$,
$C_z$ (\protect\ref{bg1a}--\protect\ref{bg1c2}) \\
\vtop{\baselineskip=6 pt \halign{#\hfil \cr
Specific \cr heat jump \cr}}& $\Delta C_{1D}$
 (\protect\ref{ac1})& $\Delta C$
 [(\protect\ref{bhk1}), Fig. \protect\ref{fig3}]\\
\vtop{\baselineskip=6 pt \halign{#\hfil \cr
Intrachain \cr coherence length \cr}}& $\xi_0$ (\protect\ref{acd1})
& $\xi_{0z}$ [(\protect\ref{bhk3}), Fig. \protect\ref{fig3}]\\
\vtop{\baselineskip=6 pt \halign{#\hfil \cr
Fluctuation  \cr region width\cr}}& $\Delta t_{1D}$
(\protect\ref{acc1})& $\Delta t_{3D}$
[(\protect\ref{bhl1}), Fig. \protect\ref{fig4}]\\
\end{tabular}
\label{table2}
\end{table}

\begin{table}
\caption{
Order of magnitude estimate of the fluctuation
parameter $\kappa$ in several materials that undergo
a three-dimensional charge-density-wave transition.
The intrachain coherence length $\xi_{0z}$, measured by
X-ray or neutron scattering, and the single-chain
coherence length $\xi_0$ are used to estimate the value
of the fluctuation parameter $\kappa$ at the three-dimensional
ordering transition (see equation (\protect\ref{bw1})).
 $\xi_0$ is estimated from
the Fermi velocity $v_F$ and the zero-temperature
energy gap $\Delta(0)$, using
equation (\protect\ref{bx1}), based on the simplest microscopic theory.
The estimates of $\xi_{0z}$, $v_F$, and $\xi_0$ are probably only
accurate to within a factor of two. Consequently, the
estimate of $\kappa$ is only accurate within an order of
magnitude. Nevertheless, clearly all these materials
lie outside the regime of validity ($\kappa < 10^{-3}$,
see Figure \protect\ref{fig2}) of  the lowest-level
 approximation (see Section \protect\ref{lla1})
and so an exact treatment of the intrachain
fluctuations of the order parameter is necessary.
Also note that these values of $\kappa$ can be used to
estimate the values of $J/|a|$ and they are in
roughly the same regime as considered in Figures 3-5.
}

\begin{tabular}{lccccc}

Material & $\xi_{0z}$ & $v_F$ & $\Delta(0)$ & $\xi_0$ & $\kappa$ \\ *[-0.05in]
& \AA & $10^{5}$ m/sec & meV & \AA & \\
\tableline
K$_{0.3}$MoO$_3$          & 15\tablenotemark[1] & 2.0\tablenotemark[2]
 & 100\tablenotemark[3] & 3 & 0.2 \\
(TaSe$_4$)$_2$I           & 60\tablenotemark[4] &  14\tablenotemark[5]
   & 200\tablenotemark[6] & 10 & 0.1 \\
K$_2$Pt(CN)$_4$Br$_{0.3}$ & 100\tablenotemark[7] & 10\tablenotemark[8]
  & 100\tablenotemark[9] & 15 & 0.1  \\
TSeF-TCNQ                 & 30\tablenotemark[10]  & 0.15\tablenotemark[10]
 & 10\tablenotemark[11]  & 2  & 0.02 \\
\end{tabular}
\tablenotetext[1]{Ref. \cite{dat}.}
\tablenotetext[2]{M. -H. Whangbo and L. F. Schneemeyer, Inorg. Chem. {\bf
25}, 2424 (1986).}
\tablenotetext[3]{L. Degiorgi, J. Phys. (France) IV
        {\bf 3}, 103 (1993).}
\tablenotetext[4]{R. Currat et al., J. Phys. (France) IV
	{\bf 3}, 209 (1993).}
\tablenotetext[5]{P. Gressier et al., Inorg. Chem. {\bf 23},
 1221 (1984).}
\tablenotetext[6]{D. Berner et al., J. Phys. (France) IV
        {\bf 3}, 255 (1993).}
\tablenotetext[7]{K. Carneiro et al., Phys. Rev. B {\bf 13}, 4258 (1976).}
\tablenotetext[8]{Ref. \cite{kim}.}
\tablenotetext[9]{P. Br\"uesch, S. Str\"assler and H. R. Zeller, Phys.
Rev. B {\bf 12}, 219 (1975).}
\tablenotetext[10]{J. P. Pouget, Semiconductors and Semimetals
  {\bf 27}, 87 (1988).}
\tablenotetext[11]{J. C. Scott, S. Etemad, and E. M. Engler, Phys. Rev. B
{\bf 17}, 2269 (1978).}
\label{table1}
\end{table}

\appendix

\section*{Previous approximate treatments of fluctuations}

\subsection{Hartree and Hartree-Fock treatment}

For completeness these approximations
which have been considered in the literature
 \cite{mas,tuc,sch} are discussed here.  The replacement
$ \mid\phi(z)\mid^4  \to  2q <\mid\phi\mid^2>  \mid\phi(z)\mid^2 $
is made in the free energy functional (\ref{aa1}) in
the partition function (\ref{af1}).
The cases $q=1$ and $q=2$ correspond to the Hartree \cite{mas} and
Hartree-Fock \cite{tuc} approximations, respectively.
 The functional integral is
then over a Gaussian field and can be performed analytically.
In effect one makes the Gaussian approximation with
\begin{equation}
a \to a + \Sigma \equiv a + 2 q  b <\mid\phi\mid^2>.
\label{av2}
\end{equation}
The expectation value $<\mid\phi\mid^2>$ is calculated self consistently
\begin{equation}
{\Sigma \over 2 q b}= <\mid\phi\mid^2>=
{1 \over 2 \pi} \int dk {T \over a + \Sigma + c k^2}
={T \over 2 \sqrt{c(a + \Sigma)}}.
\label{avv2}
\end{equation}
Consequently, the dimensionless
self energy $\tilde \Sigma \equiv \Sigma/|a|$
satisfies the self-consistent equation
\begin{equation}
\tilde \Sigma = \left({ q^2 \kappa \over 2(\tilde \Sigma -1)} \right)^{1/2}
\label{av3}
\end{equation}
Since the linear susceptibility is
\begin{equation}
\chi_1(0)={1 \over a + \Sigma}
\label{av4}
\end{equation}
the function $f(\kappa)$ in the susceptibility is given by
\begin{equation}
f(\kappa) = {1 \over 4 (\tilde \Sigma -1)}.
\label{av5}
\end{equation}
The results of the Hartree and Hartree-Fock approximations are shown
as dashed curves in Figure \ref{fig1};
they are in qualitative but not quantitative agreement
with the exact results.
As $\kappa \to 0$, $\tilde \Sigma \to 1$, and so (\ref{av3})
implies $\tilde \Sigma \to 1 + q^2 \kappa/2$. Hence, this
approximation gives
\begin{equation}
f(\kappa) \to  {1 \over 2 q^2 \kappa}  \ \ \ \  {\rm as} \ \ \   \kappa
\to 0.
\label{av6}
\end{equation}

\subsection{The lowest-level approximation
\label{lla1}}

Close to the three-dimensional
 ordering transition $\kappa \ll 1$.
This limit corresponds to the semi-classical limit of
the Schr\"odinger-type equation (\ref{ass1}).
Previously a Gaussian wavefunction, sharply peaked at
$\rho =1$, has been used as a variational ground state
wave function for
(\ref{ass1}). The results are \cite{man}
\begin{equation}
\lambda_{0,\ell}-\lambda_{0,0}=  \kappa \ell^2.
\label{au1}
\end{equation}
In this approximation the correlation length
(given by (\ref{aat1}), (\ref{an1}) and (\ref{au1})) is
\begin{equation}
\xi_1(T)={2 \mid a \mid c \over b T}.
\label{av1}
\end{equation}
This result was given previously in References \cite{sca2},
\cite{man}, and \cite{die}.
In this limit it is to be expected that $\rho^2$ has an expectation
value of 1 in the ground state.
If it is further assumed that the $n=0$ states form a complete
set (the lowest-level approximation) then
\begin{equation}
1= < u_{0,0} | \rho^2 | u_{0,0} >
= < u_{0,0} | \rho e^{-i \varphi}
\left(\sum_\ell | u_{0,\ell} >< u_{0,\ell} | \right)
\rho e^{i \varphi} | u_{0,0} >
= \mid < u_{0,1} | \rho | u_{0,0} > \mid ^2
\label{aqr2}
\end{equation}
The expression (\ref{aqq1}) then reduces to
\begin{equation}
f(\kappa) = {1 \over \kappa}.
\label{avv1}
\end{equation}
Note that in the lowest-level approximation
the sum on the left hand side of (\ref{aqr1}) is only $1/2$.

In the lowest-level approximation
the dimensionless function $g(\kappa)$ in the
third-order susceptibility is given by
\begin{equation}
g(\kappa)=
 4 {\mid <u_{0,1}|\rho|u_{0,0}> \mid ^4
\over (\lambda_{0,1} - \lambda_{0,0})^3 }
-2{ \mid <u_{0,1}|\rho|u_{0,0}> \mid ^2
\mid <u_{0,2}|\rho|u_{1,0}> \mid ^2
\over (\lambda_{0,1} - \lambda_{0,0})^2 (\lambda_{0,2} - \lambda_{0,1})}
= {7 \over 2 \kappa^3}
\label{bl2}
\end{equation}
Figure \ref{fig2} shows how the results of (\ref{avv1})
and (\ref{bl2}) deviate significantly
from the exact results for $\kappa > 10^{-3}$.
Previously Bishop and Krumhansl \cite{bis}
pointed out the shortcomings of the
lowest-level approximation for the
case of a real order parameter.

{\it The three-dimensional transition.}
Scalapino, Imry and Pincus \cite{sca2}, Manneville \cite{man},
and Dieterich \cite{die}
studied the three dimensional transition using the
lowest-level approximation
(defined in Section \ref{lla1}), i.e., they assumed that
only the lowest eigenstate (for $n=0$) for each of the
angular momentum values
$\ell = 0, \pm 1, \ {\rm and} \ \pm 2$, is important.
As mentioned earlier they also
assumed that the single chain coefficients
$ a$, $b$, and $c$ are  independent of temperature.
The zero-wavevector linear susceptibility is given by
\begin{equation}
\chi_1(0)= {4 \over \kappa \mid a \mid} =
    {2 a^2 c \over b^2 T^2}.
\label{bk1}
\end{equation}
Solving (\ref{bh1}) for the transition temperature gives
\begin{equation}
T_{3D}= \sqrt{2 J c} \ { \mid a \mid \over b}.
\label{bl1}
\end{equation}

Equation (\ref{bhk1}) for the specific heat jump reduces to
\begin{equation}
\Delta C= {16 \over 7 \xi_1}.
\label{bp1}
\end{equation}
The coherence length parallel to the chains is
\begin{equation}
\xi_{z0}= {\xi_1 \over \sqrt{2}}.
\label{br1}
\end{equation}
The transverse coherence lengths (for $J_x=J_y$) are
\begin{equation}
\xi_{x0}= {a_x \over \sqrt{8}}  \ \ \ \ \ \ \ \ \ \
\xi_{y0}= {a_y \over \sqrt{8}}
\label{bs1}
\end{equation}
The above expressions (\ref{bl1}--\ref{bs1})
were previously obtained by Manneville \cite{man} and
Dieterich \cite{die}.

In the lowest-level approximation the expression (\ref{bhl1}) for
the width of the critical region (for $J_x=J_y$) reduces to
\begin{equation}
\Delta t_{3D}= \left({7 \over 8\pi}\right)^2 \simeq 0.08.
\label{bu1}
\end{equation}


\begin {references}
\bibitem{coe}
The derivation of these coefficients is a subtle matter.
For example, for a CDW transition the presence
of a pseudogap in the electronic density of
states\cite{lee,ric,sad} and thermal lattice motion\cite{mck2}
significantly modify these coefficients.
\bibitem{kap}
If $a$ is given by (\ref{ab1}) then
$\kappa=2 t^2 ( \Delta t_{1D}/ ( 1-t))^3$
where $\Delta t_{1D}$ is given by the
one-dimensional Ginzburg criterion (\ref{acc1}).
Scalapino, Sears and Ferrell \cite{sca} assumed that $T \sim T_0$
and in their notation $\Delta t \equiv 2 \Delta t_{1D}$.
Hence  $( 1-t)/\Delta t$ in their paper corresponds
to $(1/ 4\kappa)^{1/3}$.
\end{references}
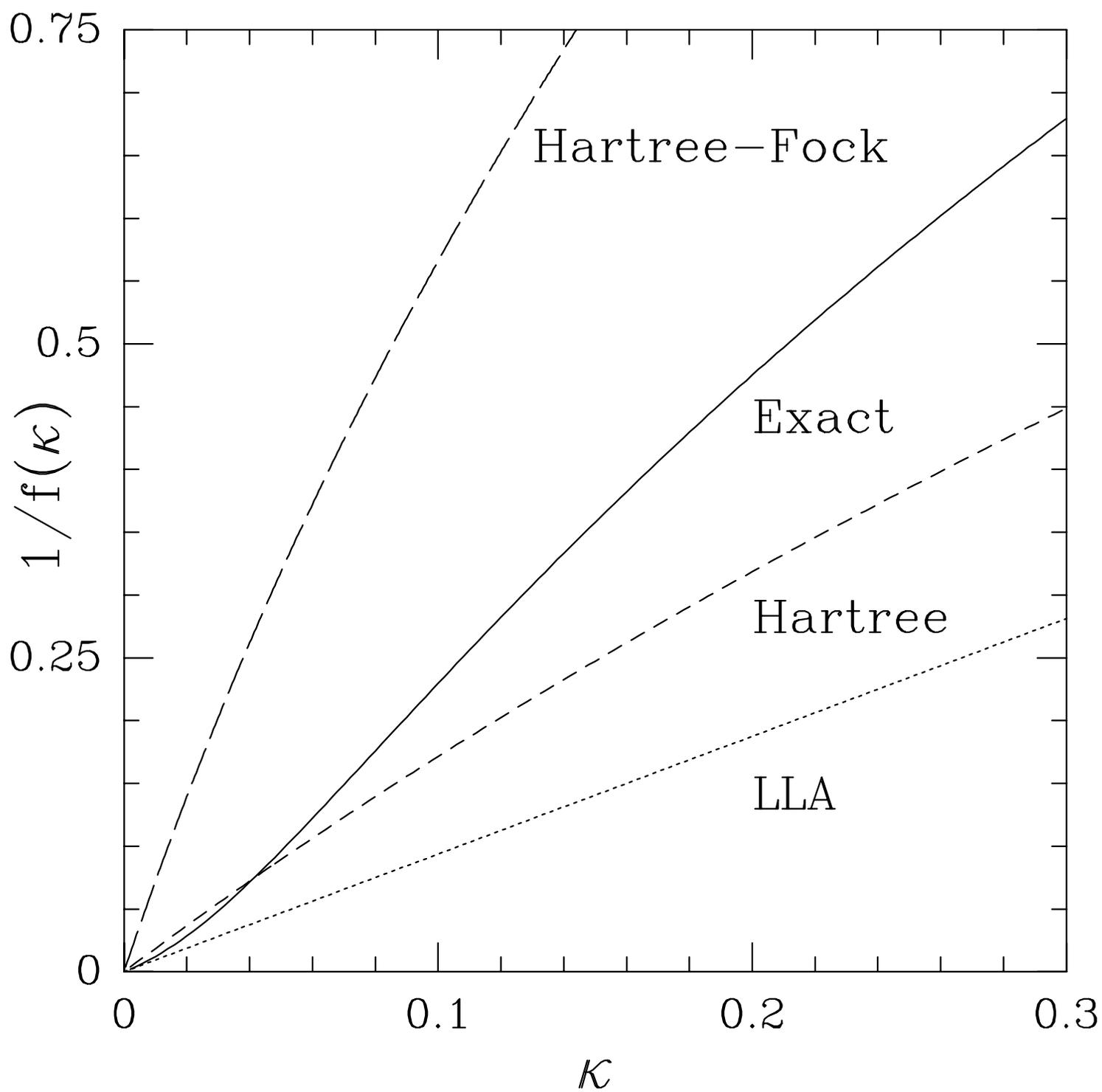
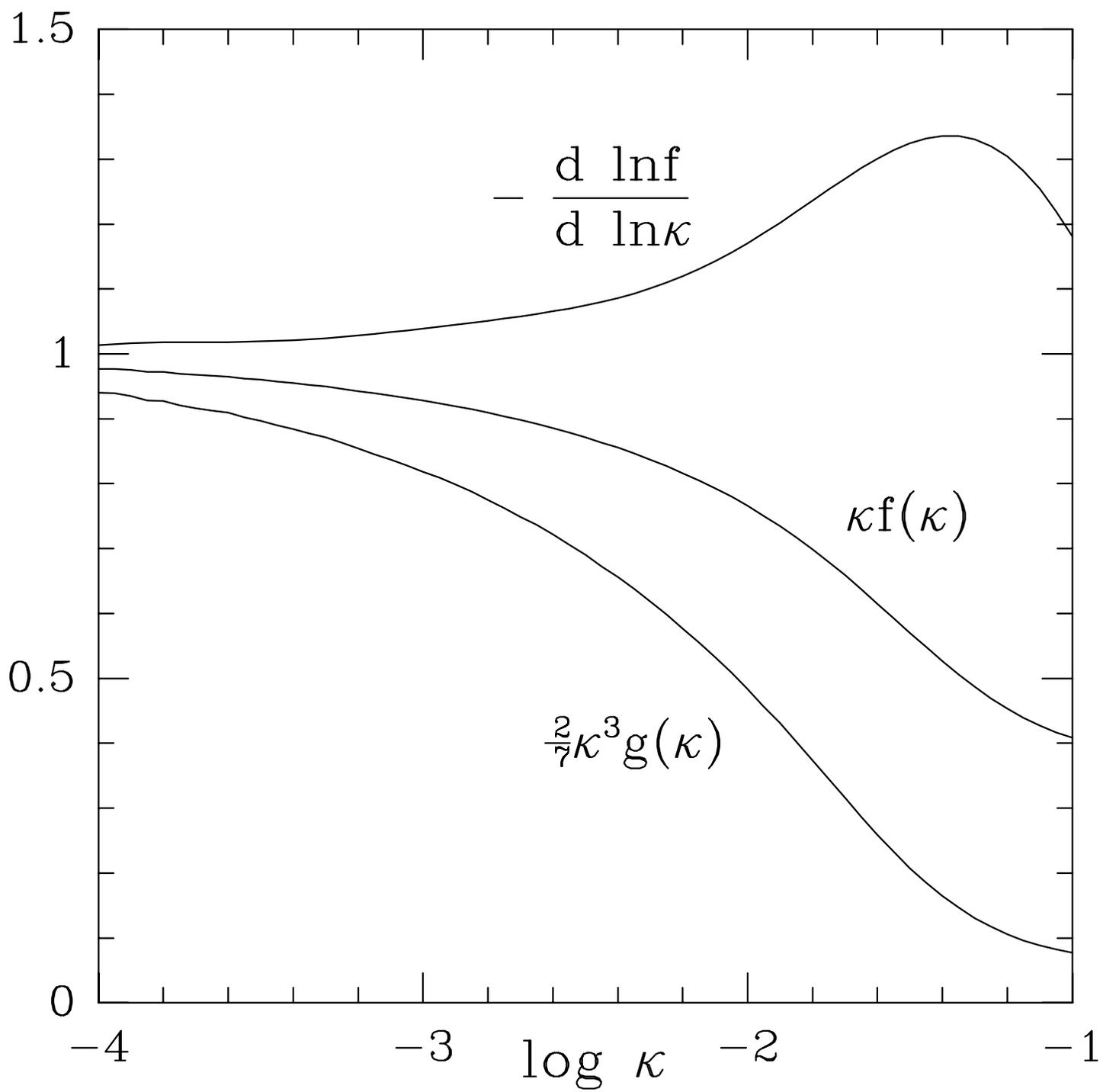
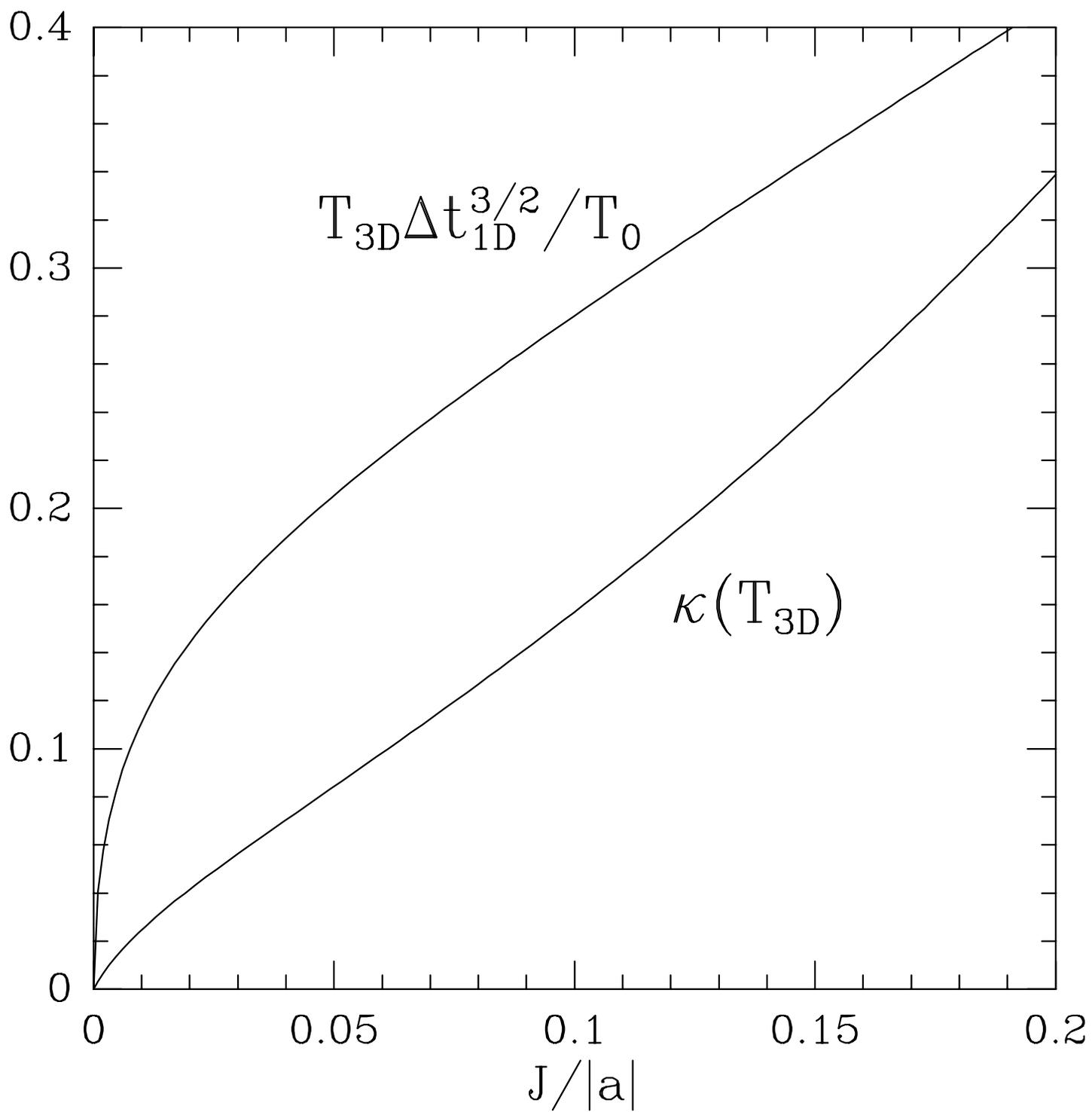
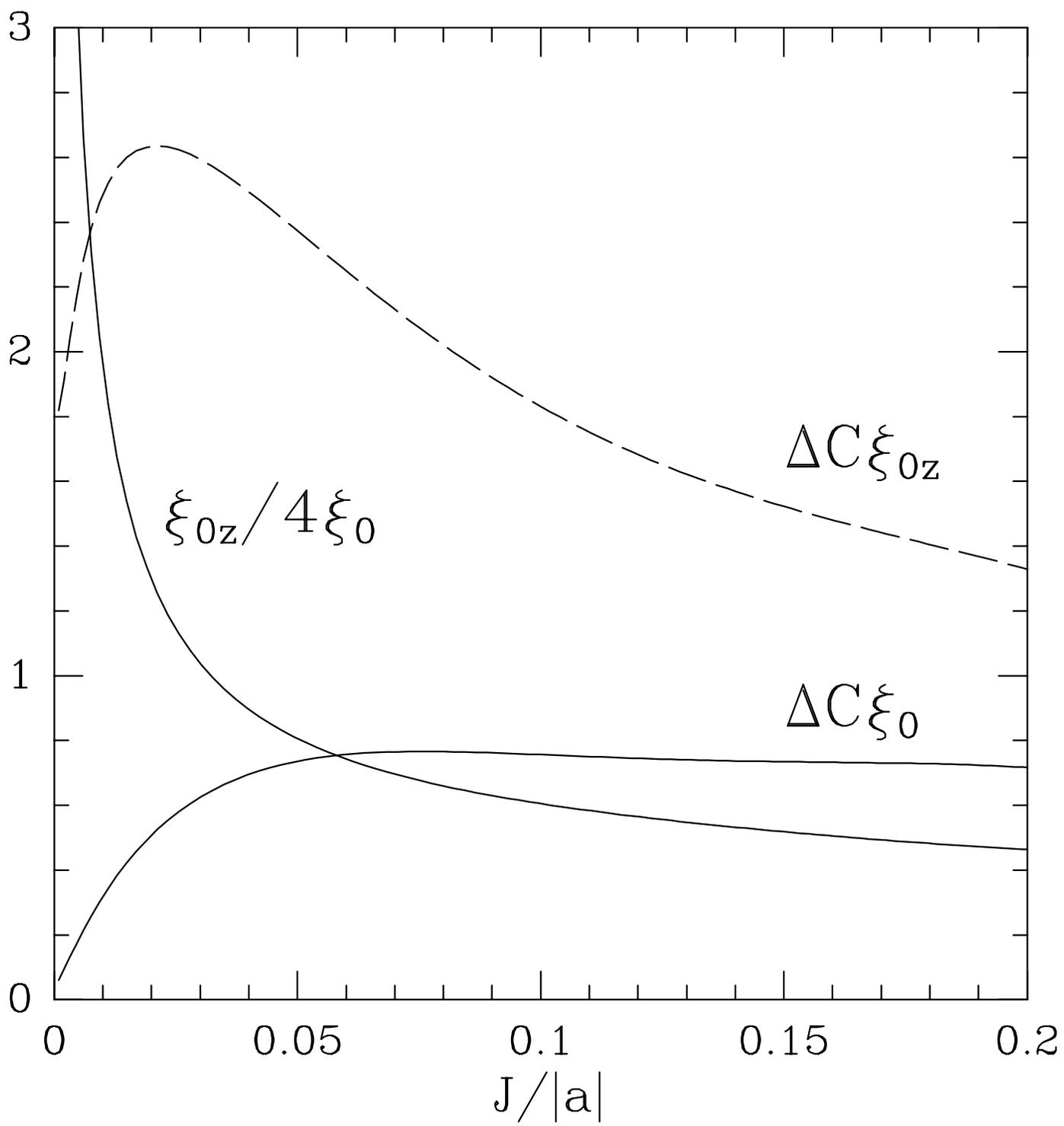
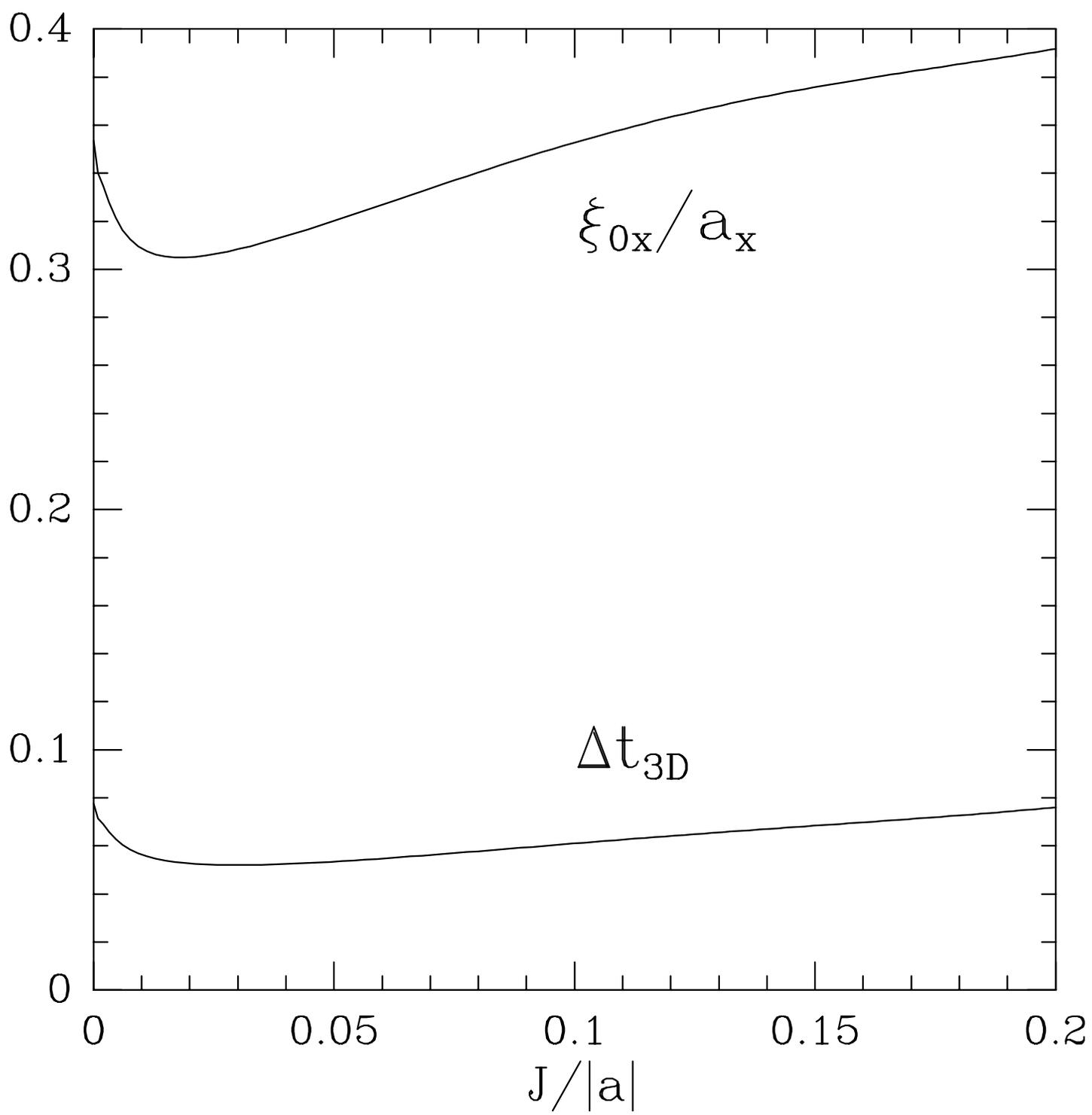
\end{document}